\documentclass[usenatbib,usegraphicx]{mn2e}
\usepackage{fleqn}
\usepackage{float}
\usepackage{color}
\usepackage{amsmath}
\usepackage{amssymb}
\usepackage{float}
\restylefloat{table}
\usepackage{array}
\usepackage{booktabs}
\usepackage{soul}%for strikeout

\def\aj{AJ}%
\def\apj{ApJ}%
\def\apjl{ApJ}%
\def\aap{A\&A}%
\def\mnras{MNRAS}%
\def\pasp{PASP}%
\def\nat{Nature}%
\title[Merger-Induced Metallicity Dilution]
{Merger-Induced Metallicity Dilution in Cosmological Galaxy Formation Simulations}

\author[S.~Bustamante et al.]
{\parbox{17cm}
{Sebasti\'{a}n Bustamante$^{1}$, Martin Sparre$^{2,3,4}$, Volker Springel$^{1,5,6}$, and\\ Robert J.~J.~Grand$^{1,5}$}\vspace*{0.3cm}\\
$^1$Heidelberger Institut f\"{u}r Theoretische Studien,
  Schloss-Wolfsbrunnenweg 35, 69118 Heidelberg, Germany\\
$^{2}$Institut f\"ur Physik und Astronomie, Universit\"at Potsdam, Karl-Liebknecht-Str.\,24/25, 14476 Golm, Germany\\
$^{3}$Leibniz-Institut f\"ur Astrophysik Potsdam (AIP), An der Sternwarte 16, 14482 Potsdam, Germany\\
$^{4}$Department of Physics, Kavli Institute for Astrophysics and Space Research, MIT, Cambridge, MA 02139, USA\\
$^5$Zentrum f\"ur Astronomie der Universit\"at Heidelberg, Astronomisches Recheninstitut, M\"{o}nchhofstr. 12-14, 69120, Heidelberg, Germany\\
$^6$Max-Planck-Institut f\"ur Astrophysik, Karl-Schwarzschild-Str. 1, 85741 Garching, Germany
}

\begin{document}
\pagerange{\pageref{firstpage}--\pageref{lastpage}}
\pubyear{2016}

\maketitle

\label{firstpage}

\begin{abstract} 
  Observational studies have revealed that galaxy pairs tend to have lower
  gas-phase metallicity than isolated galaxies. This metallicity deficiency can be
  caused by inflows of low-metallicity gas due to the tidal forces
  and gravitational torques associated with galaxy mergers, diluting
  the metal content of the central region. In this work we
  demonstrate that such metallicity dilution occurs in 
  state-of-the-art cosmological simulations of galaxy formation.
  We find that the dilution is typically 0.1 dex for major mergers,
  and is noticeable at projected separations smaller than $40$ kpc.
  For minor mergers the metallicity dilution is still present, even though the
  amplitude is significantly smaller. Consistent with previous
  analysis of observed galaxies we find that mergers are outliers from
  the \emph{fundamental metallicity relation}, with deviations being
  larger than expected for a Gaussian distribution of residuals. Our large
  sample of mergers within full cosmological simulations also makes it possible
  to estimate how the star formation rate enhancement and gas consumption timescale behave as a
  function of the merger mass ratio. We confirm that strong starbursts
  are likely to occur in major mergers, but they can also arise in
  minor mergers if more than two galaxies are participating in the
  interaction, a scenario that has largely been ignored in 
  previous work based on idealised isolated merger simulations.  
\end{abstract}

\begin{keywords}
methods: numerical -- galaxies: interactions -- galaxies: star formation -- galaxies: evolution
\end{keywords}

%==================================================================================================
\section{Introduction}
%==================================================================================================

The interstellar medium (ISM) in galaxies is steadily enriched by heavy elements formed during the life and death of stars. This enrichment is accompanied by accretion of low-metallicity gas from reservoirs surrounding galaxies. Observations yield a correlation between the stellar mass ($M_\star$) and the metallicity of the star-forming gas ($Z_\text{gas}$) of galaxies \citep{Tremonti-2004}.
The existence of this correlation shows that the heavy chemical elements of a galaxy are -- at least to first order -- gradually built up during its lifetime. However, the relation has a significant scatter of 0.1 dex
\citep{Tremonti-2004}, indicating that galaxies evolve on diverse paths through the $M_\star$--$Z_\text{gas}$-plane. 

A partial reason for the scatter in this relation is that galaxies with high star formation rates (SFR) at a fixed stellar mass usually have lower metallicities than galaxies with smaller SFR. This behaviour is encoded in the observation that galaxies evolve on a two-dimensional curved surface, dubbed the \emph{fundamental metallicity relation} (FMR), in a three-dimensional space defined by the SFR, $M_\star$ and $Z_\text{gas}$ of galaxies \citep{Ellison-2008, Mannucci-2010}. While the presence of this relation has been confirmed by several studies \citep{2010A&A...521L..53L, 2013ApJ...772..141B, 2013MNRAS.436.1130S, 2015PASJ...67..102Y}, it has, however, been questioned whether the FMR is \emph{fundamental} (i.e.~redshift-independent) because several observations at $z\sim 2.3$ reveal differences from the FMR in the local Universe (\citealt{2015ApJ...808...25S, 2016MNRAS.458.1529B,2016ApJ...817...10G,2017arXiv171100224S}, see also \citealt{2012MNRAS.427.1953C} for a study of $1<z<6$ lensed galaxies). \citet{2017arXiv171100224S} suggest that the redshift-evolution of the FMR can be explained by an evolution of the mass-loading factor at fixed stellar mass, and by an evolution of the metallicity of infalling gas. 

Even though the FMR might be redshift-dependent, it is interesting to ask what is driving its shape. A simple analytical model by \citet*{2013MNRAS.430.2891D} explains the local FMR by accounting for star formation, inflow of metal-poor gas from the intergalactic medium, and outflow of gas from the interstellar medium. The FMR has also been reproduced in more sophisticated hydrodynamical galaxy formation models \citep{2016MNRAS.459.2632D, De-Rossi-2017, 2017MNRAS.467..115D, 2017arXiv171105261T}.

Galaxy pairs are observed to have lower gas-phase metallicities and an increased SFR compared to isolated galaxies with comparable stellar masses \citep{Ellison-2008B,Scudder-2012,Cortijo-Ferrero-2017}. They also show a flattening of the standard radial metallicity gradient in the case of spiral galaxies \citep{Rupke-2010, Kewley-2010, Perez-2011}. Qualitatively, this agrees well with the FMR, but a careful statistical study of SDSS galaxies shows that the FMR has an overabundance of outliers (compared to a Gaussian distribution of residuals), which might be caused by interacting galaxies with strong merger-induced gas inflows that dilute the metallicity and enhance the SFR \citep{Gronnow-2015}. 

Given that observations can only probe a single instant in time of any individual merger, it is helpful to use simulations to establish how the metallicity dilution and SFR-enhancement in mergers occur. Remarkable insights have been gained about the metallicity evolution of mergers based on idealised simulations, where two equilibrium galaxies are set up to collide on a Keplerian orbit. \citet{Torrey-2012}, for example, established that metallicity dilution is associated with nuclear inflows, and metallicity enhancement is caused by chemical enrichment from active star formation. Even though idealised merger simulations can give important hints on how a merger makes the gas migrate inwards and subsequently causes a starburst, it is important to keep in mind that such simulations are not cosmologically self-consistent. Cooling of hot gas onto the ISM \citep{Moster-2011}, accretion of minor galaxies, and post-merger gas accretion \citep[which proves to be important, e.g.][]{Sparre-2017} are usually not included in idealised simulations. Properly accounting for these phenomena gives a more realistic view of how the metallicity dilution and SFR-enhancement occur in real galaxies.

In this paper we study the metallicity dilution of galaxies in galaxy mergers based on a large set of high-resolution cosmological simulations. In addition to having realistic collision orbits and structures of the involved galaxies, using cosmological simulations also makes it possible to self-consistently study how the galaxy properties evolve before and after the merger under the influence of the surrounding gas reservoirs. We can use this to reveal for the first time how simulated mergers evolve relative to the FMR. 

In Section~\ref{SecSample}, we present the Auriga simulations on which our analysis is based, and we briefly discuss the implemented physics model and the selection of our sample of mergers. In Section~\ref{DilutionStarbursts}, we correlate both metal dilution and SFR-enhancement with the mass-ratio of the mergers. We also study how the metallicity dilution depends on the separation of the merging galaxies. In Section~\ref{ScalingRelations} we analyse how mergers behave relative to the FMR. Finally, we discuss our results and conclude in Sections~\ref{Discussion} and \ref{Conclusion}, respectively.

%==================================================================================================
\section{Simulations and sample selection}\label{SecSample}
%==================================================================================================

%--------------------------------------------------------------------------------------------------
\subsection{Auriga simulations}
%--------------------------------------------------------------------------------------------------

The simulations studied in this paper are based on the Auriga project \citep{Grand-2017}, which comprises a set of 30 cosmological magneto-hydrodynamical zoom simulations of the formation of late-type isolated galaxies within Milky Way mass dark haloes. These haloes were extracted from a parent dark matter only simulation in a periodic cube of 100 comoving Mpc on a side from the EAGLE project \citep{Schaye-2015}, and were selected to have a viral mass\footnote{Defined to be the mass inside a sphere in which the mean matter density is 200 times the critical density, $\rho_\text{crit} = 3 H^2(z)/(8\pi G)$.} in the range $10^{12}<M_{200}/M_{\odot}< 2\times10^{12}$, and to satisfy a mild isolation criterion at $z=0$. However, no selection criteria were imposed at $z>0$, therefore the merger histories are unconstrained and exhibit a variety of evolutionary paths. The simulations were carried out with the moving-mesh code AREPO \citep{Springel-2010, Pakmor-2016}. A Planck-2014 $\Lambda\text{CDM}$ cosmology \citep{Planck2014} was adopted, specified by $\Omega_{\text{m}} = 0.307$, $\Omega_{\text{b}} = 0.048$, $\Omega_{\Lambda} = 0.693$ and Hubble constant $H_0 = 100\, h\, \text{km}\, \text{s}^{-1}$, with $h = 0.6777$. We refer the reader to \citet{Grand-2017} for a detailed and comprehensive description of the simulations.

%--------------------------------------------------------------------------------------------------
\subsection{Definition of the merger sample}
%--------------------------------------------------------------------------------------------------

To construct our global merger sample we use the 30 Auriga simulations at resolution \emph{level 4}\footnote{In the nomenclature of \citet{Grand-2017}.}. Galaxies and merger trees are identified by standard methods \citep{Rodriguez-Gomez-2015,Springel-2005}. We define merger events as times when a galaxy has more than one direct progenitor and each progenitor has a stellar mass of at least $1\%$ the stellar mass of the main progenitor. This is done in order to avoid spurious effects introduced by unresolved objects. With these conditions, we obtain 137 mergers.

For each galaxy merger we define the stellar mass-ratio, $\mu$, as the mass of the secondary galaxy divided by that of the main progenitor. We determine the stellar masses at the time when the mass of the secondary galaxy reaches it local maximum before the merger. We define the merger ratio in such a way that we always have $\mu \leq 1$.

To characterise the mergers we also calculate the \emph{time of first pericentral passage}, $t_{\text{per}}$, and the \emph{time of final coalescence}, $t_{\text{coal}}$. We define the latter as the peak time of the central black hole accretion rate, or the time at which the mean radial distance of the 10 most bound particles of the secondary progenitor reaches the half-mass radius of the main progenitor, whichever is later. As a caveat, due to the relatively sparse temporal frequency at which the snapshots of the simulations are stored, i.e. $\sim 100\ \mbox{Myr}$ at the redshift range of our sample, all temporal quantities associated to the mergers have an intrinsic uncertainty of about this value.

We restrict ourselves to coalescence times corresponding to $z_\text{coal}\leq 1.5$, and we require the most massive galaxy in a merger to have at least $M_*\geq 10^{9.5}{\rm M}_\odot$. We briefly remind the reader that the limited $z=0$ mass range of the Auriga galaxies sets an upper limit of $M_* \simeq 2\times10^{11} {\rm M}_\odot$ on our merging galaxies.

To create a master sample consisting only of \emph{clean mergers}, from our global sample we exclude those that have other mergers in a short timespan; we specifically require no other mergers with $\mu>1:100$ to occur within 1 Gyr of the first pericentral passage and of final coalescence, i.e.~in the time interval $ t_{\text{per}}-1\ \text{Gyr} < t < t_{\text{coal}}+1\ \text{Gyr} $.

In a few instances, our merger tree method has problems to properly track the main progenitor. This happens when we have two galaxies of similar mass experiencing a fly-by, but no subsequent merger. In these cases the merger tree algorithm may swap the assignment of being the main progenitor several times back and forth between the two galaxies. We exclude epochs from our sample where such progenitor swaps occur.

With these selection criteria we obtain a master sample of 70 mergers, allowing us to statistically study the properties of mergers. To summarize, all these mergers occur at $z<1.5$ and have masses of $10^{9.5}{\rm M}_\odot\leq M_* \leq 2\times10^{11} {\rm M}_\odot$. The total gas fractions of the mergers, defined relative to the total halo mass, range from $9\%$ to $16\%$, meaning that they are typically gas-rich, and thus ideal to study the process of metallicity dilution.

%..................................................................................................
%FIGURE 1:
\begin{figure*}
\centering
\includegraphics[width=0.99\textwidth]{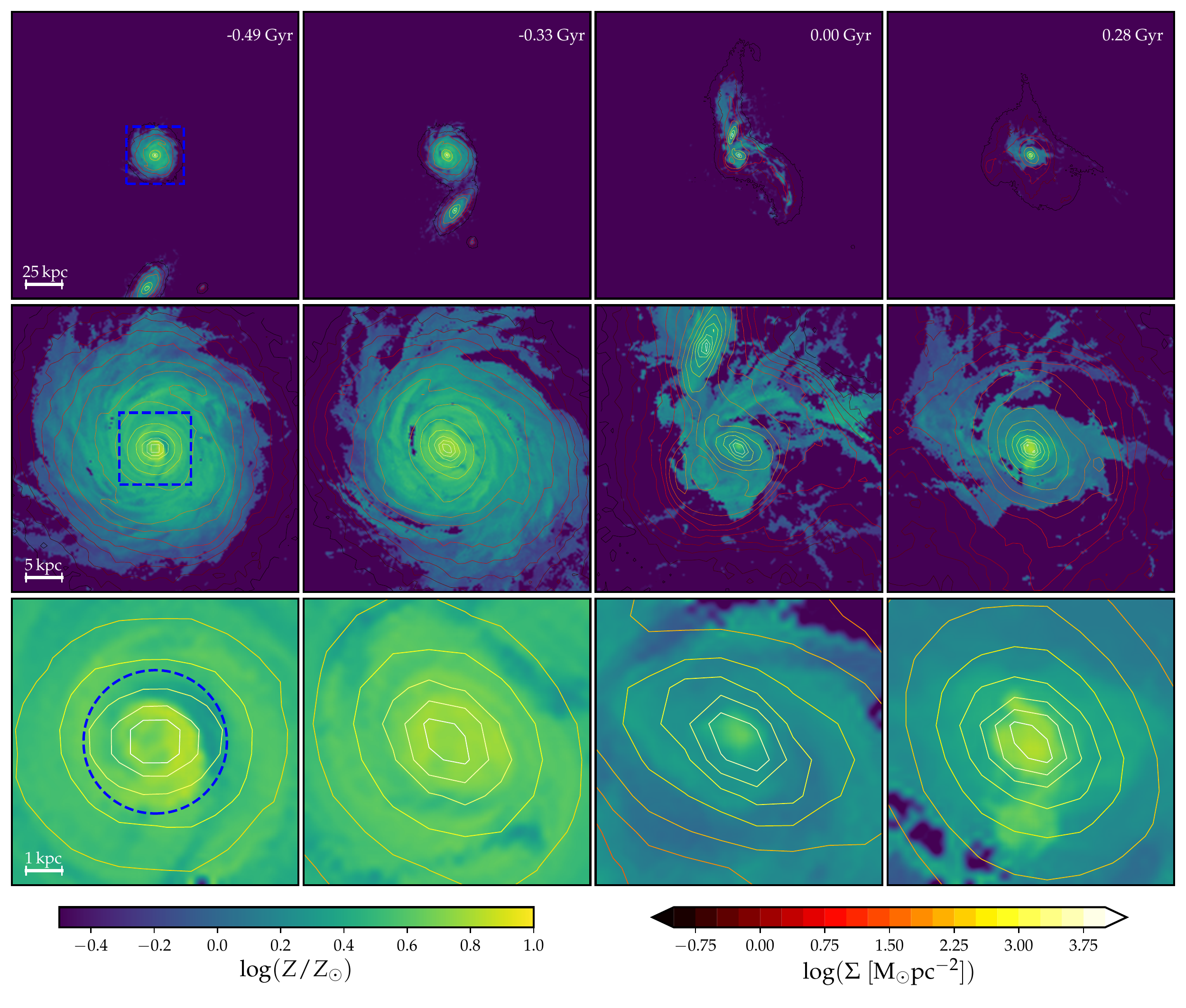}
  \caption{Projection of the SFR-weighted gas-phase metallicity (colour map) and stellar surface density (colour contours) of the primary galaxy in \emph{merger 1}. From left to right we show 4 different snapshots of the evolution, corresponding to $-0.49$, $-0.33$, $0$ and $0.28$ Gyr. From top to bottom we show different zoom levels with 200, 40 and 8 kpc on a side, respectively. For the first and second rows, the white dashed-line squares enclose the region shown in the next zoom level. For the third row, the white dashed-line circles correspond to the central 2 kpc-radius region for which we calculate all the studied properties. The metallicity in the nuclear region reaches its lowest value close to the coalescence time ($t=0$ Gyr), when the gas inflow is strong. Furthermore, we note that outside this region there is also a somewhat weaker dilution, indicating that the dilution process extends across the galactic disc. In the post-merger stage, the metallicity increases again in an inside-out way.}
  \label{fig:Metallicity_Projection}
\end{figure*}
%..................................................................................................

%--------------------------------------------------------------------------------------------------
\subsection{A sample of isolated galaxies}\label{IsolatedSample}
%--------------------------------------------------------------------------------------------------

In Section~\ref{sec:DistributionOfMergers} we will compare our merger sample with isolated galaxies. We construct a suitable comparison sample by marking those Auriga galaxies as isolated where no mergers with $\mu\geq 1:100$ occur within 1 Gyr. To be specific, we exclude all times within time-intervals,  $ t_{\text{per}}-1\ \text{Gyr} < t < t_{\text{coal}}+1\ \text{Gyr} $, of any $\mu\geq 1:100$ merger.

%--------------------------------------------------------------------------------------------------
\subsection{Physics model}
%--------------------------------------------------------------------------------------------------

All simulations use the Auriga galaxy formation model, see \citet{Grand-2017} for a comprehensive overview. As highlighted in \citet{Torrey-2012}, there are four key processes that influence the evolution of metallicity and star formation in galaxies, namely nuclear inflows of low-metallicity gas, chemical enrichment from active star forming regions, galactic outflows and locking of gas-phase metallicity in the stellar phase. Metallicity dilution and SFR-enhancement can therefore be understood as a consequence of the competition between these processes, where usually nuclear inflow and chemical enrichment dominate at low and high redshift, respectively. Below we concisely describe how these different aspects are accounted for in the Auriga physics model \citep[for more details, see][]{Vogelsberger-2013, Marinacci-2014, Grand-2016}.

For modelling the interstellar medium (ISM), the subgrid two-phase model first presented in \citet{Springel-2003} is implemented. In this, star-forming gas cells are considered to be composed of two different phases, namely a cold, dense phase embedded into a hot, diffuse ambient medium. Above a density threshold of $n = 0.13\ \text{cm}^{-3}$, gas cells are assumed to enter a thermally unstable star-forming regime where, according to the \citet{Chabrier-2003} initial mass function, they are stochastically converted either into a star particle or a SNII feedback-induced wind particle launched in an isotropic direction. In the former case, the created star particle represents a single stellar population with a given age, mass and metallicity. Metal enrichment and mass loss from SNIa \citep{Thielemann-2003, Travaglio-2004} and AGB stars \citep{Karakas-2010} are modelled by calculating how much mass per star particle moves off the main sequence at every time step. The returned metals and gas mass are then deposited into nearby gas cells with a top-hat kernel. 

Note that the implemented gas-recycling approach is drastically different from the stochastic approach presented in \citet{Torrey-2012}, where a star particle has a probability to become a SPH gas particle depending on a characteristic recycling timescale. Although their approach gives a less accurate tracking of the spatial and temporal distribution of metals and mass, it allows the tracing of the origin of any particular mass or metal element, thereby allowing to differentiate enriched metals from an initially set metallicity profile. In our case of cosmological simulations, all the metal content is produced from star formation enrichment within the simulation. The wind particles are launched with a metal content determined by the initial gas-phase metallicity, and are re-coupled to the gas once they reach a gas cell with a density below $0.05$ times the density threshold for star formation or a maximum travel time is exceeded. In either of these cases, their metal, mass, momentum and energy contents are deposited into the local gas cell. This ensures that the wind material leaves its launching site, favouring the production of gas outflows in the simulation.

Additionally, we include prescriptions for a uniform background UV field for reionization (completed at $z=6$), primordial and metal line cooling, magnetic fields, as well as subgrid models for black hole seeding, growth through accretion and associated feedback.

%==================================================================================================
\section{Metallicity dilution and starburst properties}\label{DilutionStarbursts}
%==================================================================================================

To introduce the concept of metallicity dilution we show the evolution of the SFR-weighted gas-phase metallicity of the primary galaxy of one of the major mergers in our sample in Figure~\ref{fig:Metallicity_Projection}. Note that we use SFR-averaged metallicities to mimic observations of H\textsc{ii} regions of star-forming gas. The projections in Figure~\ref{fig:Metallicity_Projection} are done onto the galactic plane for 3 different zoom levels of 200, 40 and 8 kpc. In the analysis we adopt a solar metallicity of $Z_{\odot} = 0.02$. For the first two snapshots, shown in the first and second columns, we see that the metallcity profile does not change appreciably, and it is only when the secondary galaxy comes closer than $40$ kpc that the metallicity gets significantly diluted, as seen at the coalescence time in the third column (see also \emph{merger 1} in Figure~\ref{fig:PropertiesMassiveMergers}). In the fourth column, 0.28 Gyr after the merger, the enhanced SFR replenishes the metal content of the gas in the central region through stellar enrichment, defining thus the end of the dilution period. In the following, we study the dilution process in galaxy mergers in more detail.

%--------------------------------------------------------------------------------------------------
\subsection{Time evolution}\label{time-evolution}
%--------------------------------------------------------------------------------------------------

%..................................................................................................
%FIGURE 2:
\begin{figure*}
\centering
  \includegraphics[width=1.0\textwidth]{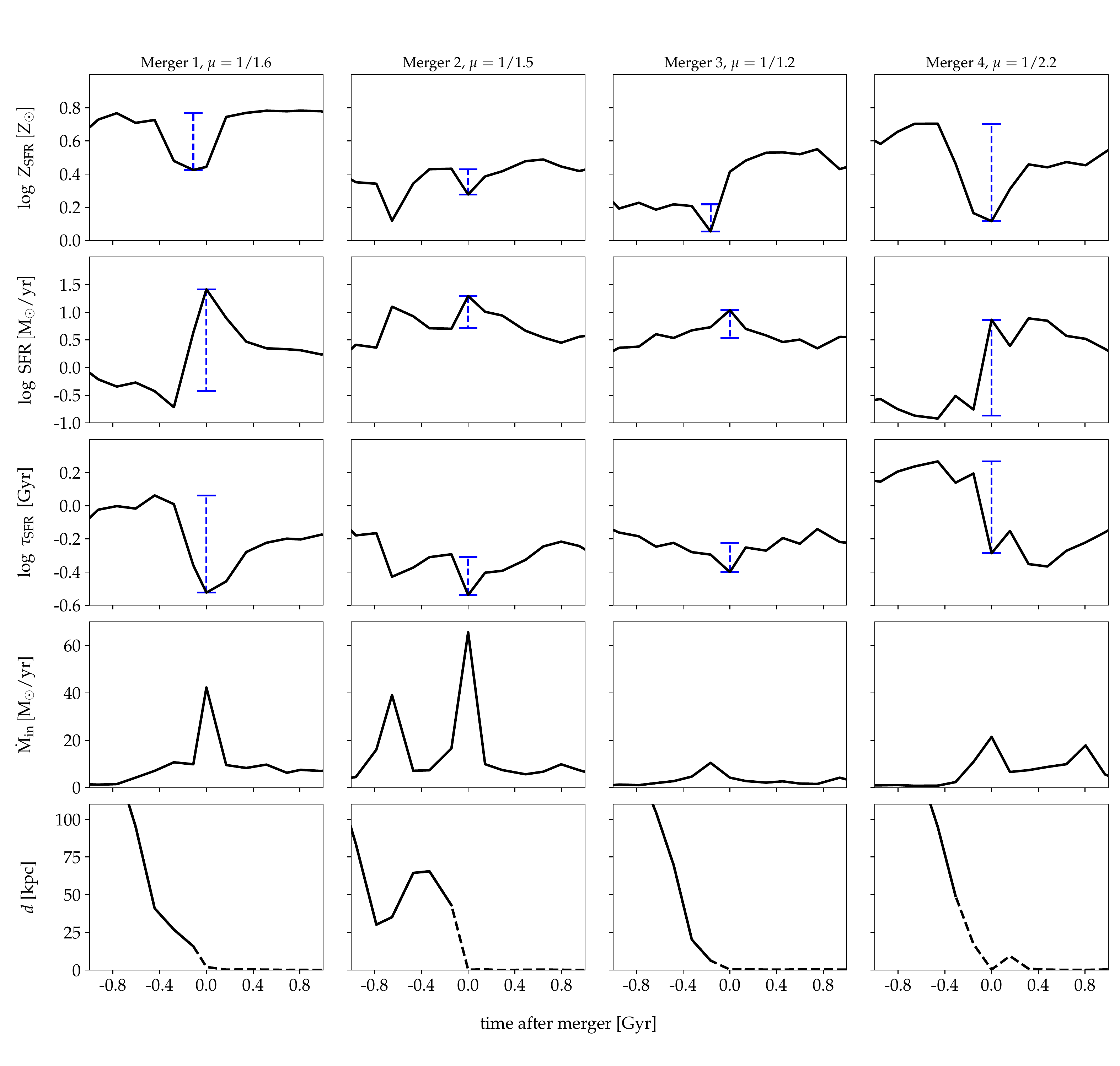}
  \caption{We show four typical major mergers, each presented in a separate column headed by the mass ratio. From top to bottom we show the time-evolution of the gas metallicity inside 2 kpc (first row), the SFR (second row), the gas consumption timescale (third row) and the nuclear gas inflow rate (fourth row). We also show the radial distance between the two galaxies (fifth row), where dashed lines correspond to a reconstruction of the orbit based on the most bounded particles of the secondary galaxy after both galaxies have fused in the merger trees. The $x$-axis shows the time elapsed since $t_{\text{coal}}$, with negative values corresponding to pre-coalescence stages. In the first three rows, vertical dashed lines correspond to the fractional changes of the respective quantities. We see that these major mergers of our sample cause a significant metallicity dilution (especially for \emph{merger 1} and \emph{merger 4}), strong SFR-peaks, and a decrease in the gas consumption timescale, demonstrating that those mergers form stars in a more bursty mode.}
  \label{fig:PropertiesMassiveMergers}
\end{figure*}
%..................................................................................................

To introduce the primary quantities we use in our analysis, we plot the evolution of four different mergers from our sample in Figure~\ref{fig:PropertiesMassiveMergers}. These mergers are selected as representative cases for major mergers (defined to have $\mu > 1/3$, which is also the same criterion used to select major mergers, e.g., in \citealt{Lotz-2008}).

In order to quantify the behaviour of metallicity dilution and SFR-enhancement during a merger, we analyse the nuclear region of the main progenitor. We define the nuclear region to be a sphere of radius $2\ \text{kpc}$ around the galactic centre (identified with the minimum of the galactic gravitational potential). This cut is chosen to mimic the effect of a limited aperture of a SDSS fibre. For each snapshot we define the SFR-averaged metallicity as
\begin{equation}
\label{eqn:Z_SFR}
Z_\text{gas} = \frac{ \sum_i \text{SFR}_i  Z_i }{\sum_i \text{SFR}_i} ,
\end{equation}
where $i$ runs over all gas cells inside the nuclear region, $\text{SFR}_i$ and $Z_i$ are the instantaneous star formation rate and the fraction of gas mass in metals in the $i$-th gas cell, respectively.

In the first row in Figure~\ref{fig:PropertiesMassiveMergers}, the gas-phase metallicity has been calculated based on the gas cells inside the galaxy corresponding to the main branch of the merger tree, as a function of time. The evolution of $Z_\text{gas}$ reveals that these galaxies experience a dilution of metallicity right before or at the time of the SFR peak. To quantify the metallicity dilution, we define the pre-merger background metallicity level as the previous local maximum prior to the time when the metallicity minimum occurs. We refer to the time where we determine the background metallicity as $t_{Z_\text{max}}$. In the plot, the background and minimum metallicity levels are marked as the ends of the dashed vertical lines, which represent the metallicity dilution times. Based on the ratio of these two levels we define $\Delta \log Z_\text{gas}$, which we will refer to as the metallicity dilution.

The second row shows the SFR, which is calculated in the same fashion as the metallicity. The SFR peaks, marked by dashed vertical lines, are determined as the largest SFR-value of the main progenitor within 1 Gyr of the merging time registered in our merger trees. To determine the fractional increase in SFR due to the merger-induced starbursts, we measure the pre-merger background SFR-level by applying a smoothing algorithm to filter out short timescale fluctuations over 1 Gyr. Then, we find the local minimum before the SFR-peak. The background SFR-level is determined as the corresponding value in the original non-smoothed profile, and the time at which this occurs defines $t_{\text{SFR}_\text{min}}$. The fractional SFR-enhancement is then calculated as the ratio between the peak SFR and the background level. In the remaining parts of the paper we will use $\Delta \log \text{SFR}\equiv \log [\text{SFR}(\text{peak}) / \text{SFR}(\text{background})]$ to quantify this fractional enhancement.

We see that the two strongest fractional metallicity dilutions occur in \emph{merger 1} and \emph{merger 4}. For \emph{merger 2} and \emph{merger 3}, we see a somewhat weaker dilution. Establishing the presence of metallicity dilution in cosmological simulations is a notable result, since this has so far mainly been studied in idealised merger simulations. The only other work that has found this effect in a cosmological simulation is \citealt{2017arXiv171105261T} (see the left panel of their Fig. 9), who identified a merger-induced epoch of metallicity dilution in a $M_* = 10^9 M_\odot$ galaxy.

The third rows show the ISM gas consumption timescale, calculated as the ratio of the ISM gas mass ($M_\text{ISM}$, calculated using Eqn.~1 from \citealt{Torrey-2012b}) and the SFR of the galaxy. We calculate this based on the ISM-gas-cells of the primary galaxy. The evolution of the gas consumption timescale reveals that merger-induced starbursts are typically associated with a decrease in the ISM gas consumption timescale, quantified in a similar fashion as for the gas-phase metallicity dilution, i.e.~as the logarithm of the ratio between the nearest local minimum to the SFR-peak and the pre-merger local maximum $\Delta \log \tau_{\text{SFR}}$. This behaviour is also identified in the merger simulations of \citet{Sparre-2016}, which uses the same physical galaxy formation model as the present paper. A decrease in $\tau_{\text{SFR}}$ shows that star formation occurs in a more bursty mode than for normal star-forming galaxies. A similar bursty mode is observed in galaxy mergers and ULIRGS \citep{Sanders-1991, Daddi-2010, Krumholz-2012, Scoville-2013}.

Additionally, we show the nuclear gas inflow rate and relative radial distance (fourth  and fifth rows, respectively). Note that all SFR-peaks and metallicity dilutions are accompanied by a conspicuous increase in the nuclear gas inflow, supporting thus the scenario where infalling pristine gas disrupted during the merger is the main driving mechanism of the dilution. The post-merger gas-phase metallicity level can vary widely from galaxy to galaxy depending on the infalling gas supply. For example, for \emph{merger 1} and \emph{merger 2}, the post-merger levels are similar to their pre-merger values owing to comparable pre- and post-merger gas supplies. In \emph{merger 3}, the gas supply is not enough to counteract enrichment from recently formed stars, and the post-merger metallicity is thus much larger than the pre-merger value. The opposite case occurs in \emph{merger 4}, where a continuous post-merger gas supply keeps diluting the metallicity. 

%\todo{MS: compare with Fig 9 of \citet{2017arXiv171105261T}}

%--------------------------------------------------------------------------------------------------
\subsection{Correlations}
%--------------------------------------------------------------------------------------------------

%..................................................................................................
%FIGURE 3:
\begin{figure}
\centering \includegraphics[width=0.30\textheight]{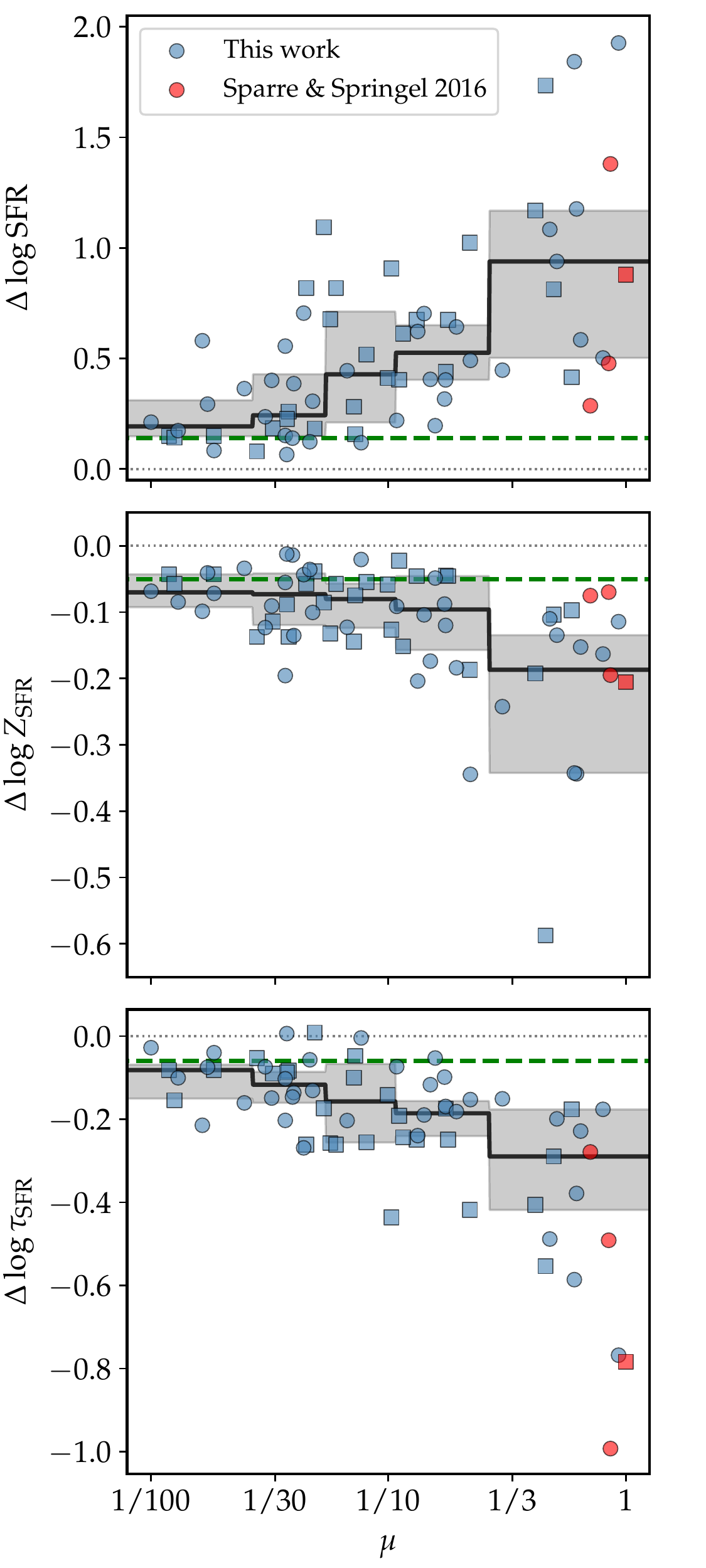}
  \caption{Dependence of the fractional enhancement/dilution of SFR (top panel), metallicity (middle panel) and gas depletion timescale (bottom panel) on the merger mass-ratio ($\mu$). Solid lines correspond to median values, and grey regions to 25\%-75\% percentiles. The squares show galaxies where the merging pair is gravitationally affected by a third galaxy nearby (see text for details), and the circles show mergers where no other galaxies are nearby. All the properties exhibit a gradually increasing dependence on merger ratio. If we only take into account galaxies without a nearby third galaxy these results are consistent with previous idealised simulations of mergers, where $\mu > 1/3$ mergers can cause strong starbursts and $\mu<1/3$ mergers cause more modest starbursts. A full inclusion of the cosmological environment, such as in the form of additional galaxies in the vicinity of the merger, can, however, significantly increase the SFR and decrease the gas depletion timescale of minor mergers.}
  \label{fig:Correlations}
\end{figure}
%..................................................................................................

Having shown that four of our mergers in subsection~\ref{time-evolution} have a strong SFR-peak associated with a decrease in the gas consumption timescale and the nuclear metallicity, we will now study the behaviour of these quantities in our full merger sample. In order to do so, we show how $\Delta \log \text{SFR}$, $\Delta \log Z_\text{gas}$ and $\Delta \log \tau_\text{SFR}$ correlate with $\mu$ in Figure~\ref{fig:Correlations}.

First, we see that the strongest SFR-enhancements occur in mergers with large merger ratios. For example, the three highest values of $\Delta \log \text{SFR}$ occur in galaxies with $\mu>0.5$. This is in good agreement with other studies showing that major mergers cause strong starbursts \citep{Mihos-1996, 2004MNRAS.350..798B, Springel-2005b, Hayward-2014}. Specifically, the range of our relative SFR-enhancements for major mergers is in good agreement with what is typically found in idealised simulations \citep{2008MNRAS.385L..38M,2008A&A...492...31D,2010ApJ...720L.149T,2013MNRAS.430.1901H,2015MNRAS.452.2984K, 2015MNRAS.448.1107M,2016A&A...592A..62G}.

For low merger mass-ratios of $1/30<\mu<1/3$ there are a few mergers with relatively large SFR-enhancement of around 1.0 dex. Further inspection shows that while the merger goes on, there is another galaxy on its first passage around the main galaxy of the merger. The time between coalescence of the two galaxies, which are about to merge with the main galaxy, is more than 1 Gyr, which is why these galaxies have not been filtered out by our selection criteria for mergers to be \emph{clean}. We have marked all these merging systems, which are affected by a third galaxy, with a square in the figure. The presence of systems where more than two galaxies are participating in the merger is a clear consequence of our use of cosmological simulations, where mergers do not occur in isolation. An interesting conclusion is that a minor merger -- if we include the full cosmological structure in the surroundings including other galaxies -- can trigger a star formation enhancement which is comparable to a major merger. This is not the first time such systems of multiple interacting galaxies are reported. For example, one of the systems (1349-3) from the simulations of \citet{Sparre-2016} also consists of multiple merging galaxies, and the observations of Stephan's Quintet also reveal such a system. Even though such systems of multiple interacting galaxies are expected to occur in the real Universe, they have, however, been largely overlooked in previous generation's of idealised modelling of merging galaxies.

On average, minor mergers with $\mu \simeq 0.1$ cause substantial SFR-enhancements of about 0.5 dex (a factor of 3). For lower merger ratios of $\mu<0.05$, the SFR-enhancements are more modest, in most cases lower than a factor of 2. In idealised simulations, \citet{Cox-2008} found the same conclusions; their merger with $\mu = 0.02$ causes an almost negligible SFR-peak, whereas their $\mu = 0.1 $ merger causes a more visible -- but still small -- SFR-enhancement.

For the metallicity, the strength of the dilution increases gradually until $\mu\sim 1$, where the peak dilution is around $0.17$ dex. We thus confirm the idealised simulations from \citet{Torrey-2012}, which also show that metallicity dilution occurs in the centre of galaxies. The magnitude of the dilution in our work is slightly larger than the maximum dilution of $\sim 0.05$ dex observed in \citet{Scudder-2012} for galaxies being $\sim 11$ kpc apart. However, there are several possible explanations for this discrepancy: first, their selection criteria for galaxy pairs do not guarantee that these systems are strictly interacting galaxies; therefore, when non-interacting systems are included, the real extent of the metallicity dilution would be likely underestimated. Second, we are reporting values near to or at the coalescence time, when the dilution is supposedly strongest\footnote{Major mergers seem to be an exception to this. See Figure~\ref{fig:Trajectory_Metallicity}. }. This time corresponds to small projected distances $\lesssim 5$ kpc, which are more challenging to probe observationally. Finally, due to their weaker dilutions, they do not split the metallicity offsets into different merger ratio bins, even including merger ratios above one, i.e.~they also use the secondary galaxies. When we do the same, we obtain a more modest median metallicity dilution of 0.08 dex, which is in better agreement. Overall, establishing the presence of metallicity dilution in cosmological simulations is an interesting finding as it shows that tidal forces and gravity torques exerted during more realistic mergers are mechanisms capable of driving enough nuclear gas inflows to produce metallicity dilution and SFR-enhancement, even in minor mergers, albeit at a lesser extent.

Epochs of very strong gas compression mostly appear when the merger ratios are high, see the median curve in the $(\mu , \Delta \log \tau_\text{SFR})$-plot. A similar bursty mode for major mergers is also found in the hydrodynamical simulations of \citet{Renaud-2014} and \citet{Sparre-2016}.

The same plot also shows that $\mu \simeq 0.3$ is a characteristic scale above which the average $\tau_\text{SFR}$-value experiences a decrease of $\sim 0.3$ dex (a factor of 2) during a merger. Moreover, four of the mergers in the same range ($\sim 30\%$) exhibit even larger decreases of $0.5$ dex up to $0.8$ dex. Therefore, using $\mu \simeq 0.3$ as a threshold to distinguish major and minor mergers captures whether bursty epochs with short $\tau_\text{SFR}$-values are likely to occur.

Finally, we quantify the fluctuations of SFR, metallicity and depletion timescale in isolated galaxies. In order to do so, we have implemented the same algorithm used for finding the peaks and dips associated to SFR-enhancements and metallicity dilution in mergers. For example, in the case of SFR, we find all the local maxima of a galaxy whenever it is marked as isolated according to our merger selection criteria. Then, for each maximum, we find the local minimum that occurs right before the time of the maximum. With these two values we quantify the fractional change of the fluctuation. For metallicity and depletion timescale, we apply an analogous procedure. We show in every panel the mean value of the fractional change of these fluctuations as green dashed lines. Mergers with small stellar mass ratios (i.e. $\mu<1/30$) exhibit fractional changes comparable to the fluctuation levels of isolated galaxies, which is consistent with the idea of very minor mergers not driving gas inward.

%- - - - - - - - - - - - - - - - - - - - - - - - - - - - - - - - - - - - - - - - - - - - - - - - - 
\subsection{Projected distance}
%- - - - - - - - - - - - - - - - - - - - - - - - - - - - - - - - - - - - - - - - - - - - - - - - - 

%..................................................................................................
%FIGURE 4:
\begin{figure}
\centering
\includegraphics[width=0.31\textheight]{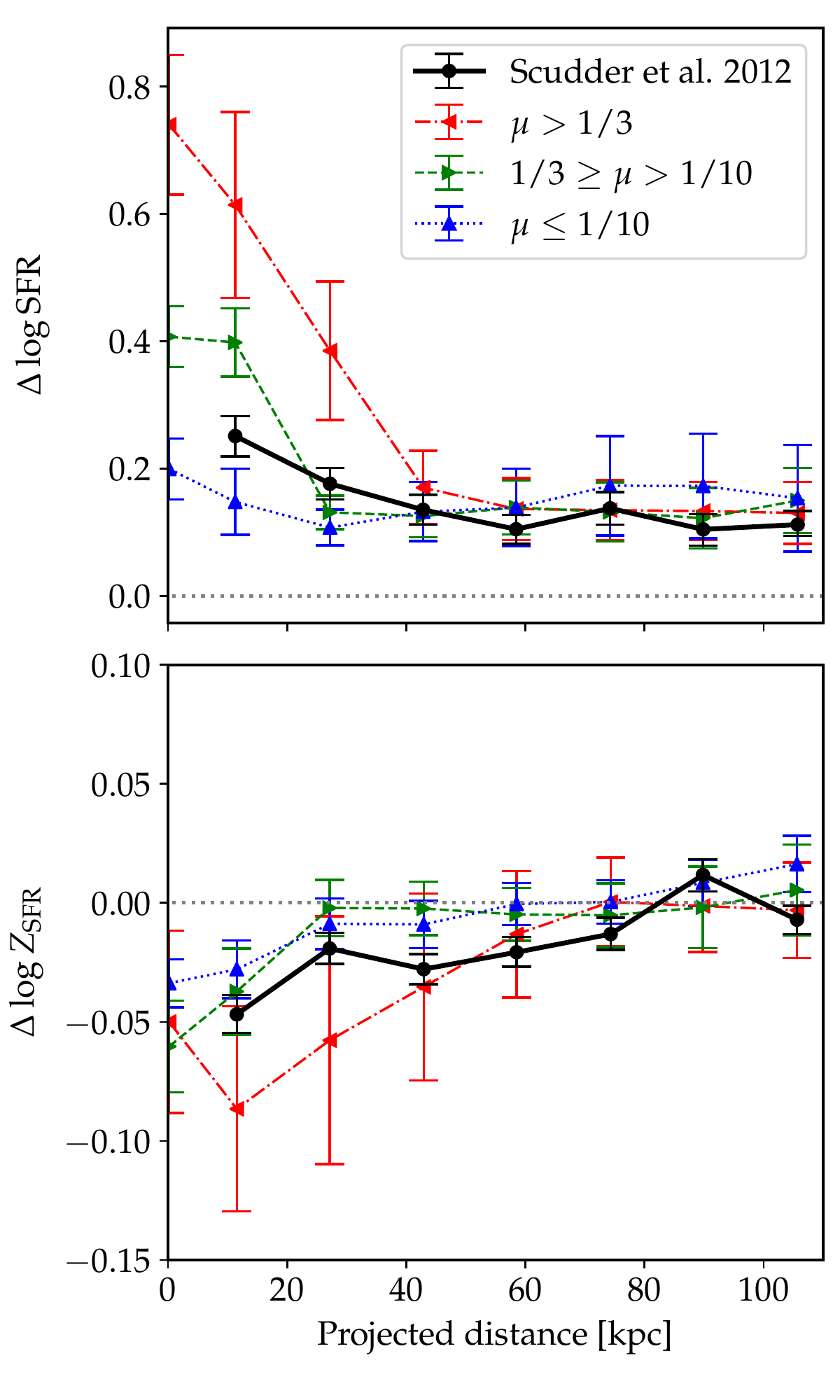}
  \caption{Fractional change of the SFR (top panel) and the metallicity (bottom panel) as a function of projected distance for our sample of galaxies (dashed lines) and the observations of \citealt{Scudder-2012} (solid lines). The error bars show the standard error of the median. The observed sample shows SFR-enhancement at $r<40$ kpc and metallicity dilution at $r<60$ kpc. For our simulated sample, major mergers show SFR-enhancement at $r<30$ kpc and dilution at $r<40$ kpc. For minor mergers, SFR-enhancement and dilution occur at the same distances, within $r<30$ kpc. The fractional changes are progressively stronger for larger mass ratios and smaller distances, with the exception of major mergers, where the dilution peak occurs at a distance $\sim 10$ kpc, which is slightly increasing with the coalescence time. This is caused by stellar metal enrichment overcoming the dilution.}
  \label{fig:Trajectory_Metallicity}
\end{figure}
%..................................................................................................

In Figure~\ref{fig:Trajectory_Metallicity} we have determined the SFR, the metallicity and the projected distance as the merging galaxies from our sample approach each other for three different mass ratio bins. The projected distance corresponds to the projection on the XY plane of the merger orbit, which means that we assume only one viewing direction per merger. Metallicity dilutions and SFR-enhancements are measured relative to the pre-merger values at $t_{Z_\text{max}}$ and $t_{\text{SFR}_\text{min}}$, respectively. Since we only have snapshots of our galaxies at a relatively sparse temporal frequency, it is not always the case that each galaxy merger is represented in every radial bin. In such a case, these galaxy mergers get a vanishing weight in those bins. Furthermore, a few mergers exhibit more complex orbits, causing their trajectories to be counted more than once in some radial bins. We evenly distribute the weights of multiple points of a given galaxy within a radial bin such that the sum of the weights is one. The final median profile is obtained by stacking all the trajectories with the proper weights in every radial bin. Besides, we include data points at $r=0$ corresponding to values at the coalescence time. Finally, to simplify comparison with the results reported by \citet{Scudder-2012}, our error bars also correspond to the standard error of the median.

The fractional changes near coalescence are stronger for larger mass ratios. For major mergers, the SFR-enhancement becomes significant at projected distances $r<30$ kpc, whereas the metallicity dilution becomes noticeable at slightly larger distances $r<40$ kpc. This difference is caused by merger-induced gas inflows diluting firstly the metals, and soon after producing a strong SFR-enhancement as a result of gas compression in the galactic centre. The magnitude of this process is also revealed by the slight increase of the metallicity at the coalescence time, caused by stellar metal enrichment starting to overcome the dilution. For minor mergers, both effects become significant at comparatively smaller distances, for $r<20$ kpc. A smaller distance of influence is expected from the modest gravitational influence of smaller galaxies. Due to weaker gas inflows in this case, the SFR-enhancement and the metallicity dilution sync up, with stellar metal enrichment overcoming the dilution only after coalescence.

Finally, we compare with the observations of \citet{Scudder-2012}. At large projected distances the SFR-enhancements exhibit a similar behaviour, although their metallicity dilution is lower than ours. The reason for the differences is probably the different adopted normalisations. While they use a carefully selected control sample from which the fractional changes are calculated, we use pre-merger values due to our limited number of galaxies. At smaller distances, their values sit between our major and minor mergers, consistent with the mixed merger ratios in the sample that they use. Taking these considerations into account we regard our simulations as consistent with the \citet{Scudder-2012} observations. Our results also agree with the idealised simulations of \citet{2013MNRAS.433L..59P}, who reproduced the observed trend from \citet{Scudder-2012}.

%==================================================================================================
\section{Mergers and the Fundamental Metallicity relation}\label{ScalingRelations}
%==================================================================================================

%..................................................................................................
%FIGURE 5:
\begin{figure*}
\centering
\includegraphics[width=0.31\textheight]{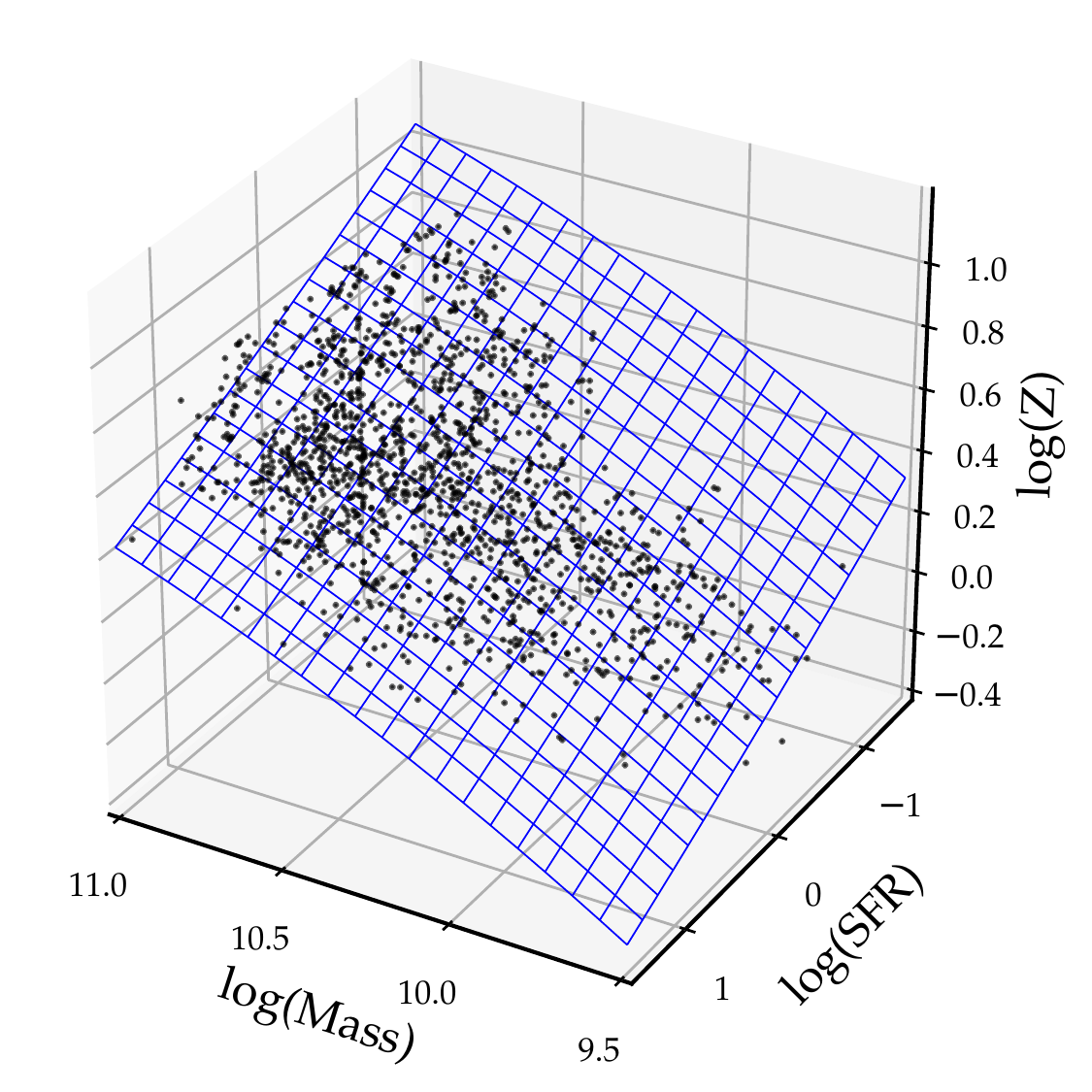}
\includegraphics[width=0.31\textheight]{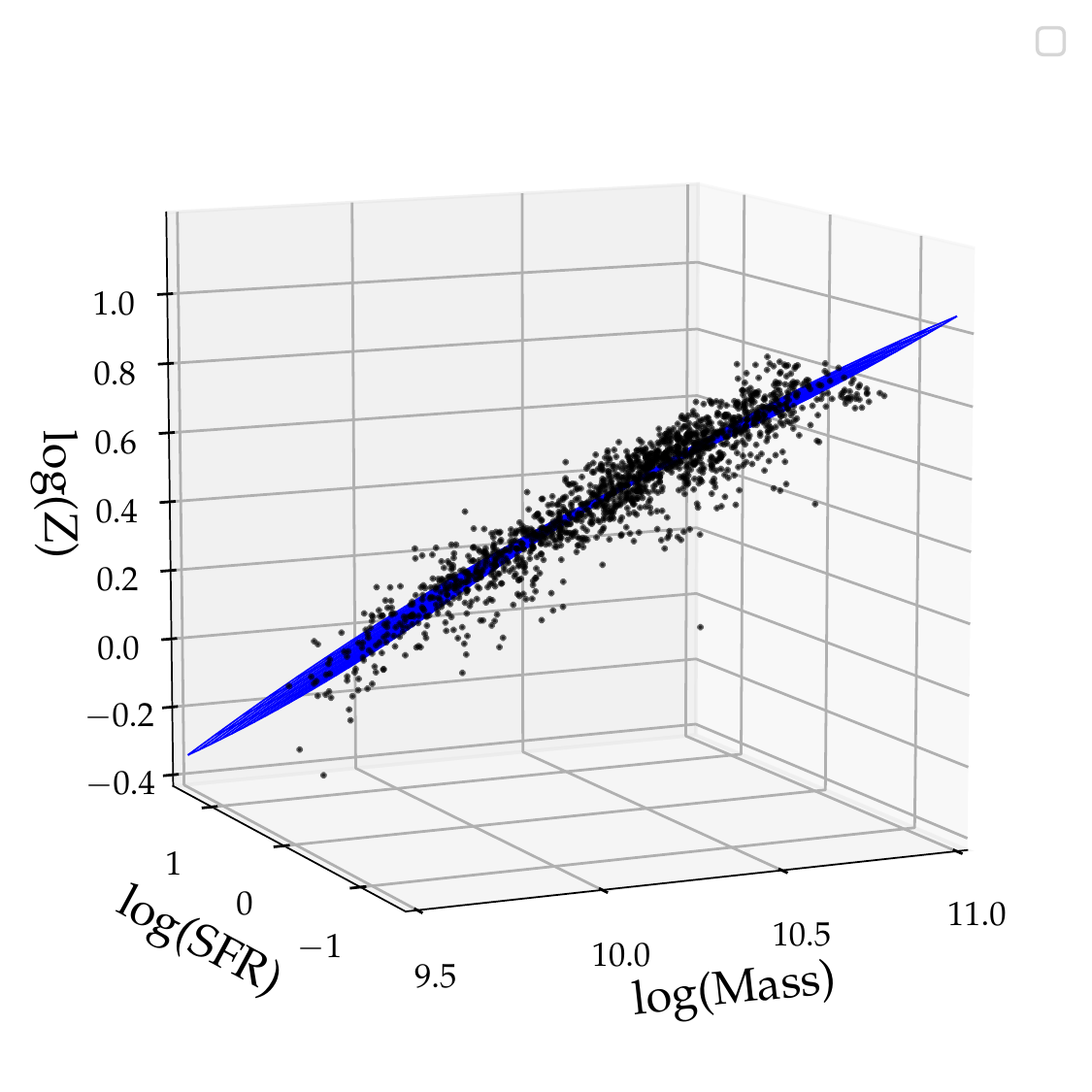}
\includegraphics[width=0.5\textheight]{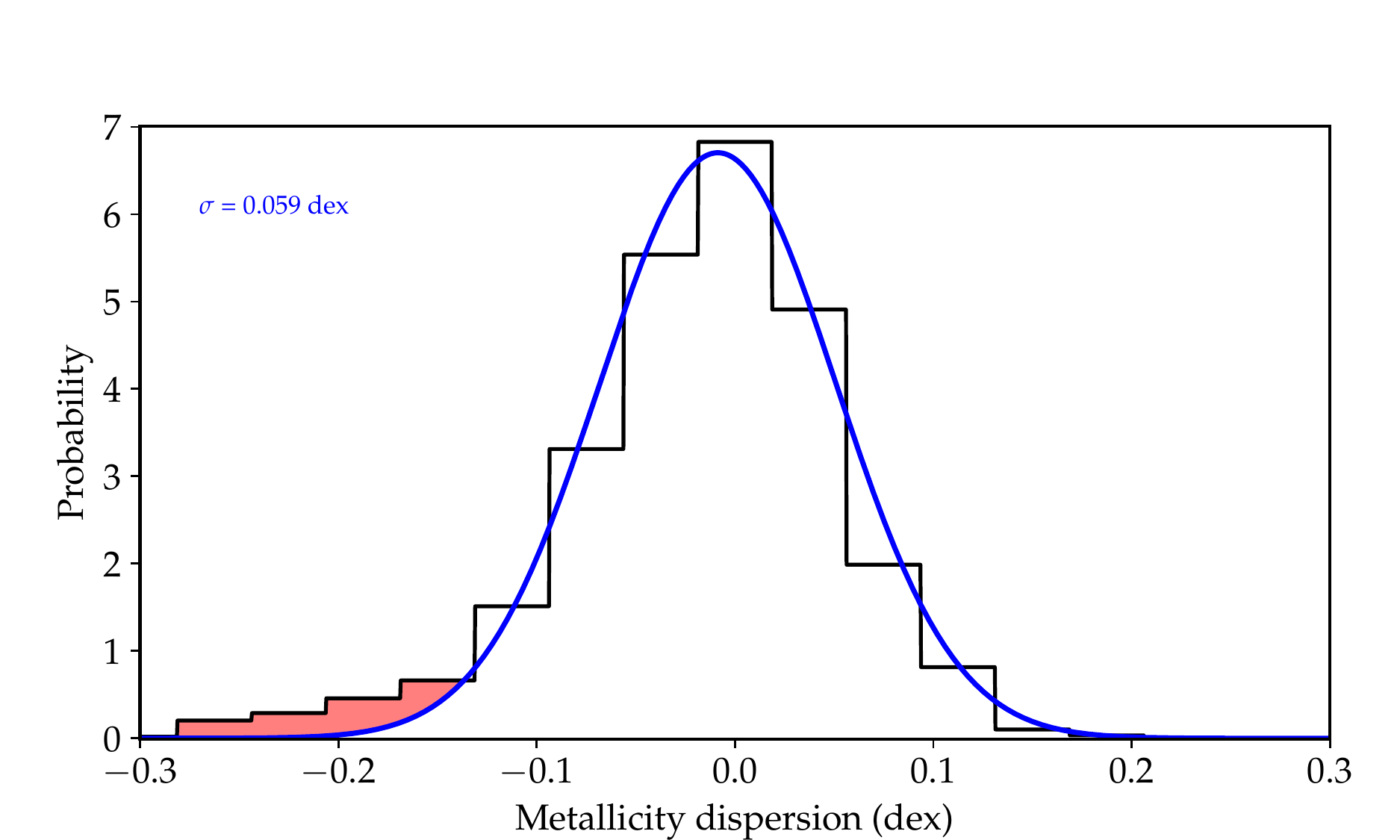}
  \caption{\emph{Top panels:} two projections of the distribution of simulated galaxies at different times in the three-dimensional space of stellar mass--SFR--metallicity. The blue wireframe shows a second-order polynomial fitted to the data. \emph{Bottom panel:} Normalised histogram of residuals around the fitted surface. We highlight the overabundance of low-metallicity outliers with a red shaded region. The blue line corresponds to a fitted Gaussian distribution function. The dispersion is similar to the value $\sigma = 0.053$ dex found by \citet{Mannucci-2010} for their galaxy sample, which indicates a similar level of consistency of our simulated galaxies with a FMR. }
  \label{fig:FMR}
\end{figure*}
%..................................................................................................

In this section we characterise how mergers are positioned relative to the fundamental metallicity relation (FMR). In order to do so, we show in the left panel of Figure~\ref{fig:FMR} the three-dimensional distribution of 1577 data points representing our galaxies in both isolated and merging phases. For each simulation we study all galaxies in 60 snapshots logarithmically distributed in the redshift range $0\leq z \leq 1.5$. The blue wireframe corresponds to a second-order polynomial in $M_\star$ and SFR fitted to the metallicity values, yielding:

%................................................
%EQUATION 2
\begin{eqnarray}
\nonumber
\log (Z_\text{gas})_\text{FMR} &=& 0.471 + 0.549\,m - 0.214\,s - 0.099\,m^2 \\
\label{eqn:fit_FMR}
&& + 0.010\,m\,s + 0.007\,s^2,
\end{eqnarray}
%................................................
where $m\equiv \log (M_\star) - 10$ and $s \equiv \log(\text{SFR})$, with $Z_\text{gas}$, $M_\star$ and SFR in units of $Z_\odot$, $M_\odot$ and $M_\odot\,\text{yr}^{-1}$, respectively. With the aim of testing the goodness of fit of this model, in the bottom panel of Figure~\ref{fig:FMR} we calculate the histogram of residuals $r\equiv \log (Z_\text{gas})_\text{data} - \log (Z_\text{gas})_\text{FMR}$, using 16 bins with a width of 0.04 dex. The metallicity dispersion is almost a Gaussian distribution, but with a noticeable tail towards lower metallicity values, as also noted in \citet{Mannucci-2010} and \citet{Gronnow-2015}. We fit a Gaussian distribution function using least squares, which yields an offset $\mu = 0.009$ dex and a standard deviation $\sigma = 0.059$ dex, which is very similar to the observed values of $\sigma = 0.053$ dex and $\sigma = 0.048$ dex for SDSS galaxies found by \citet{Mannucci-2010} and \citet{Gronnow-2015}, respectively. This shows that both observed and simulated galaxies exhibit a similar level of consistency with the evolution encoded in the  FMR.

%--------------------------------------------------------------------------------------------------
\subsection{Distribution of mergers}\label{sec:DistributionOfMergers}
%--------------------------------------------------------------------------------------------------

%..................................................................................................
%FIGURE 6:
\begin{figure*}
\centering
\includegraphics[width=0.31\textheight]{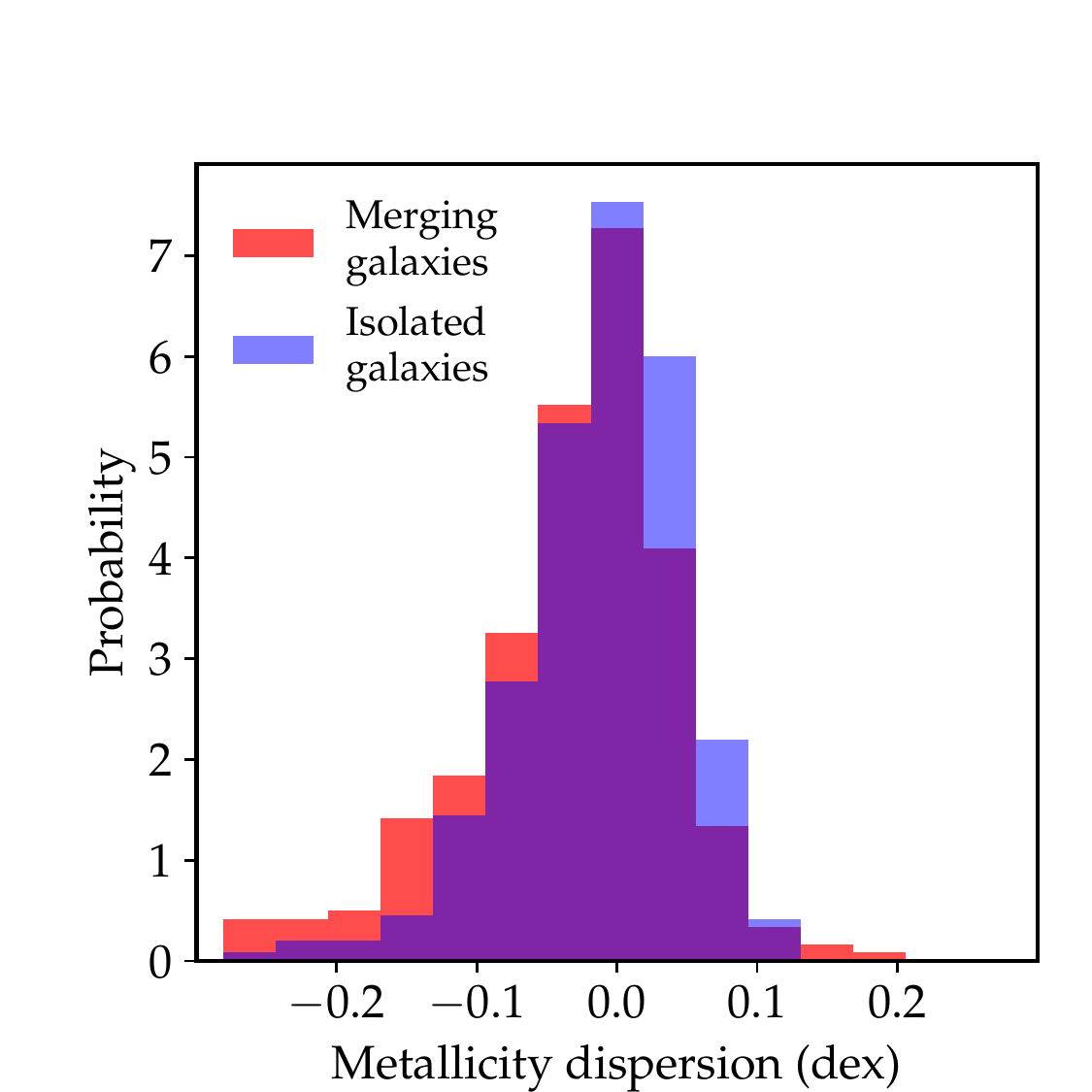}
\includegraphics[width=0.31\textheight]{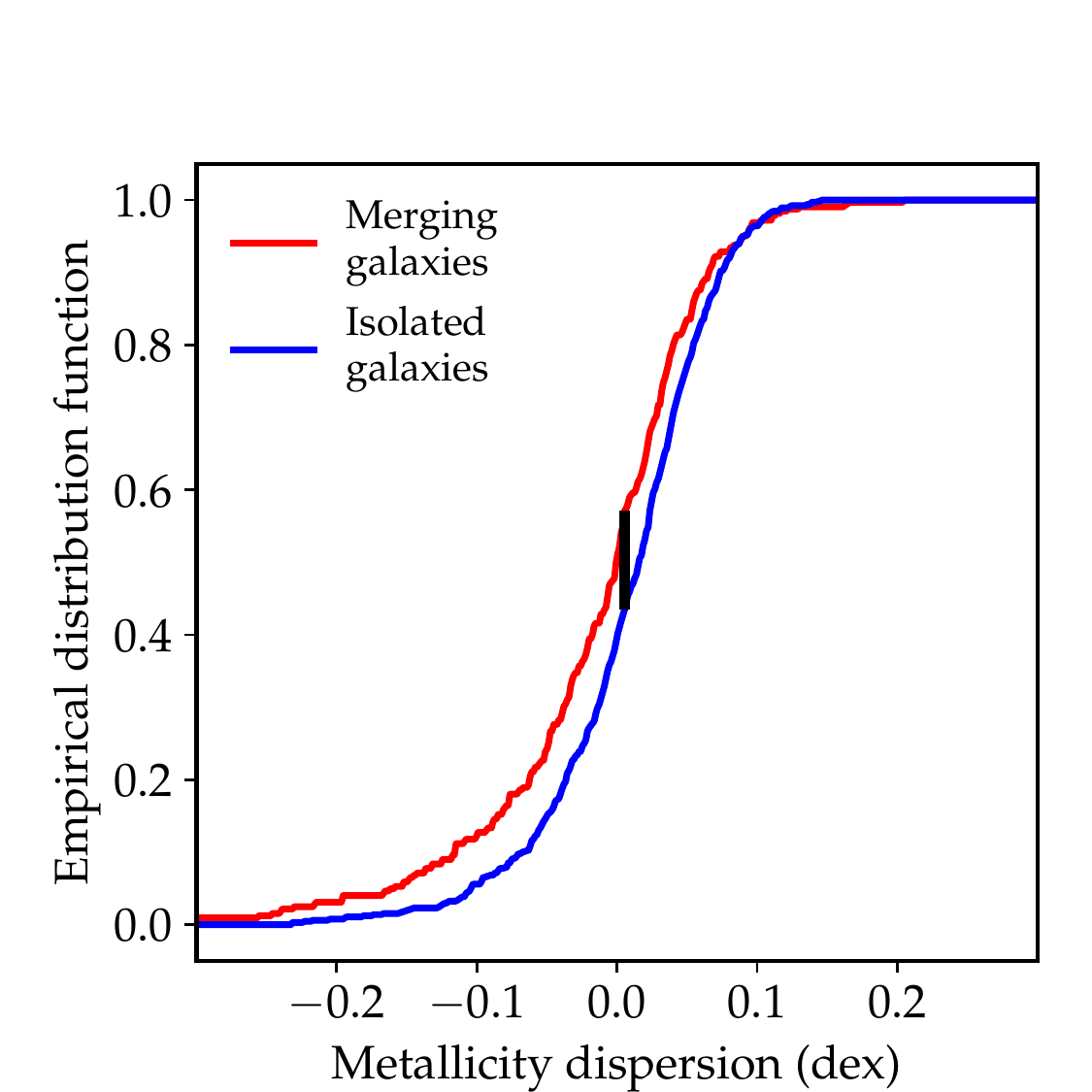}
  \caption{\emph{Left panel:} Normalised distributions of residuals of isolated galaxies (blue) and merging galaxies (red) with respect to the FMR. Merging galaxies have a more distinct tail towards low metallicities than isolated galaxies. \emph{Right panel:} Empirical cumulative distribution functions for isolated and merging galaxies. Applying a Kolmogorov-Smirnov test, both distributions differ at a confidence level of $\alpha=0.1\%$, which is good enough to affirm that isolated and merging galaxies follow different evolutionary trends. }
  \label{fig:Residuals}
\end{figure*}
%..................................................................................................

\citet{Mannucci-2010} proposed that strong low-metallicity gas inflows triggered during a merger are a plausible scenario to explain the tail of low-metallicity galaxies in the metallicity distribution. More recently, \citet{Gronnow-2015} proposed a sophisticated model to infer different properties of mergers -- merger ratios, metallicity dilutions and characteristic timescales -- from the distribution of residuals and the overabundance of outliers, showing also consistency with this scenario.

In order to study how merging galaxies are distributed with respect to the FMR, we opt to follow a different approach. Instead of trying to directly quantify the overabundance of outliers, which strongly depends on the number of merging systems included in the sample, we calculate and compare the normalised distributions of residuals for isolated and merging galaxies. To do so the isolated phases of a galaxy are determined according to the sample in Section~\ref{IsolatedSample}.

In the left panel of Figure~\ref{fig:Residuals} we show the normalised distribution of residuals. First, we note that merging galaxies exhibit a slightly higher probability to have lower than average metallicity values compared with their isolated counterparts. Moreover, they show a more distinct tail in the low-metallicity end, consistent with a merger-induced overabundance of outliers in the global distribution of residuals. The distribution of isolated galaxies also exhibits a somewhat smaller tail, which might be due to numerical issues related to our isolation criterion, or have a more physical origin; namely very minor mergers ($\mu<1/100$), long post-merger transients, or cold gas accretion.

Finally, we test how different both distributions are. We compute the empirical distribution function (ECDF) in the right panel of Figure~\ref{fig:Residuals}. Then, after applying a Kolmogorov-Smirnov test, we conclude that the null hypothesis of both distributions being numerical realisations drawn from the same underlying distribution can be rejected at a confidence level of $\alpha=0.1\%$. This is good enough to affirm that the distributions are different and that isolated and merging galaxies follow different evolutionary paths in the SFR, $M_\star$ and $Z_\text{gas}$ space. We thus confirm the result from \citet{Gronnow-2015} that mergers cause outliers in the FMR.

%--------------------------------------------------------------------------------------------------
\subsection{Evolution of mergers: stacked profiles}
%--------------------------------------------------------------------------------------------------

%..................................................................................................
%FIGURE 7:
\begin{figure*}
\centering
\includegraphics[width=1.0\textwidth]{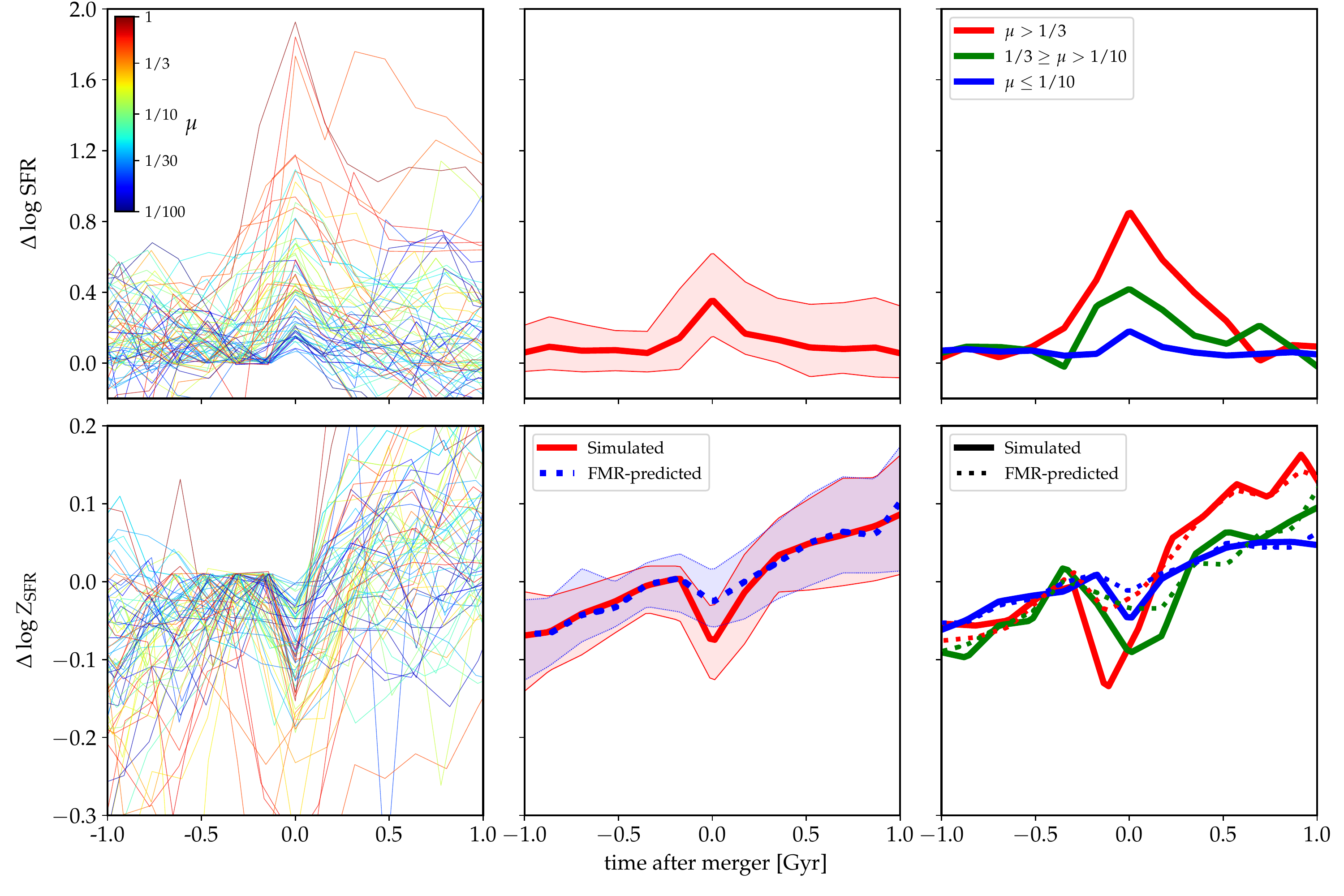}
  \caption{Fractional changes of the SFR (top-left panel) and metallicity (bottom-left panel) for the merger sample. Each merger is normalised to its pre-merger values at $t_{\text{SFR}_\text{min}}$ and $t_{Z_\text{max}}$, respectively. The $x$-axis shows the elapsed time since coalescence, with negative values corresponding to pre-merger stages. In the central panels we show the median of the stacked profiles, where the shaded regions correspond to $25\%-75\%$ percentiles. In the right panels we show the median fractional changes for 3 different mass ratio bins, namely major (red), intermediate (green) and minor mergers (blue). For the metallicity, we compute the simulated and the FMR-predicted values. The latter is computed from the stellar mass and the SFR of each galaxy along with the fitted FMR of equation~(\ref{eqn:fit_FMR}). Close to coalescence, we note that the simulated metallicity is always lower than the FMR prediction.}
  \label{fig:SFR_ZRelations}
\end{figure*}
%..................................................................................................

After establishing that merging galaxies represent a distinct population with respect to the FMR, we explore how they are evolving out of and back into the isolated phases. In Figure~\ref{fig:SFR_ZRelations}, we calculate the evolution of the fractional change of the SFR (top panels) and metallicity (bottom panels) for our merger sample. Each merger is normalised to its pre-merger values at $t_{\text{SFR}_\text{min}}$ and $t_{Z_\text{max}}$, respectively. The reference time is set to the coalescence time, with negative values corresponding to pre-merger stages.

In the central panels we show the median of the stacked profiles, where the shaded regions represent $25\%-75\%$ percentile intervals. In the top-central panel we note that galaxies in a merger start experiencing a significant SFR-enhancement at around $-0.3$ Gyr, reaching a peak of 0.36 dex at coalescence and then decreasing again to pre-merger values. A closer inspection in the top-right panel, where we use three different mass ratio bins corresponding to major, intermediate and minor mergers, shows that this evolution is mass-ratio dependent. For example, the SFR-peak values are 0.79 dex, 0.54 dex and 0.18 dex for each respective bin, and they all occur at coalescence time. This result is expected given the correlations presented in Figure~\ref{fig:Correlations}, and also applies to the strongest metallicity dilutions seen in the bottom-central panel.

A second interesting aspect is the characteristic timescale of the enhancement. Major mergers start the SFR-enhancement period at around $-0.3$ Gyr up to 0.7 Gyr, i.e.~over a timescale of 1 Gyr. For intermediate and minor mergers, this period goes from $-0.3$ Gyr  to 0.5 Gyr, and from $-0.2$ Gyr up to 0.2 Gyr, respectively.

Now we calculate the distributions of the fractional change for the simulated and the FMR-predicted metallicities. The latter is computed from the stellar mass and the SFR of each merger along with the FMR of equation~(\ref{eqn:fit_FMR}). In the bottom-central panel we see that overall the metallicity is increasing in time due to the secular evolution of the galaxy. An exception to this occurs around the coalescence time, where significant dilutions of 0.08 dex and 0.025 dex are registered in the simulated and the FMR-predicted distributions, respectively. The FMR-predicted decrease is caused because the merging SFR-enhanced galaxies naturally yield lower metallicities in the FMR relation. However, this dilution is smaller than the simulated value, thereby confirming once more that merging galaxies consist of a differentiated population.

It is interesting to note that when we look at both distributions at more distant times from coalescence, i.e.~for pre- and post-merger stages, they are more similar, meaning that the evolution is again well described by the FMR. This has the interesting consequence that a direct comparison can hint when the onset and the end of the dilution take place. In the bottom-right panel we show the dilution for the different mass ratio bins. For major mergers, we find that the onset of the dilution occurs at $-0.4$ Gyr, i.e.~0.1 Gyr before the onset of the SFR-enhancement. A time delay of 0.15~Gyr is also seen between the strongest metallicity dilution and the SFR-peak. In general, this is consistent with the results presented in Figure~\ref{fig:Trajectory_Metallicity} for metallicity versus distance projection, for which the dilution starts to be significant at slightly larger distances than the SFR-enhancement. The dilution ends shortly after coalescence at 0.2 Gyr, while the SFR-enhancement is still significantly high. About 0.5-0.7 Gyr after the end of the dilution, the SFR reaches pre-merger values and the enhancement is over. During this period of time, the stellar enrichment is able to compensate the  inflow of low-metallicity gas, placing the metallicity again close to the FMR-prediction.

For smaller merger ratios the metallicity dilution occurs at nearly similar times as the SFR-enhancement, with both having almost the same timescales \citep[see also][]{2017arXiv171111039T}. This implies that the stellar enrichment from SF activity is less efficient at compensating gas inflows than in major mergers, and the dilution only ends when the infalling gas supply is dimmed after the merger process is completed.

Metallicity dilution and SFR-enhancement timescales have been previously reported in hydrodynamical simulations of merging galaxies by \citet{Montuori-2010}. They study 1:1 gas rich mergers and find a timescale of $~1.5$ Gyr for both processes. Our SFR-enhancement timescale of 1 Gyr for major mergers is consistent with their result, although our dilution timescale is considerably shorter. A plausible explanation for this discrepancy is that their galaxies are more gas-rich, with gas fractions of 20\% compared to $9-16\%$ for our galaxies. This, along with the higher mass ratio, might cause significantly stronger low-metal gas inflows in their simulations, hence the stellar enrichment during the SFR-enhancement cannot overcome the dilution, resulting in both processes having similar timescales.

%==================================================================================================
\section{Discussion}\label{Discussion}
%==================================================================================================

Based on idealised merger simulations where two composite model galaxies collide on a Keplerian orbit, it has previously been suggested that \emph{wet mergers} can be sensibly divided into two categories; minor and major mergers. The merger mass ratios defining these two classes are typically taken to be $1/10< \mu <1/3$ and $\mu > 1/3$, respectively. The former mass-ratio reflects the limit below which a negligible SFR-enhancement arises, and the latter reflects the threshold above which a strong SFR-peak occurs. Furthermore, it has been established that major mergers also consume their gas on shorter timescales than minor mergers \citep[e.g.][]{Cox-2008}. In our cosmological simulations we confirm that stronger SFR-peaks and shorter gas consumption timescales occur for major mergers compared to minor mergers. We additionally show that a strong metallicity dilution occurs for major mergers, whereas for minor mergers the effect is more modest. We advocate that merger-induced metallicity dilution is an important addition to the standard picture of the behaviour of wet mergers. Aside from  our own simulations, this is also supported by those of \citet{Torrey-2012} and \citet{2017arXiv171105261T}. Furthermore, this effect is seen in the observations of \citet{Scudder-2012}.

Observers have for more than half a decade advocated the existence of the FMR, and recently several hydrodynamical galaxy formation simulations have reproduced this relation \citep{2016MNRAS.459.2632D,De-Rossi-2017,2017MNRAS.467..115D,2017arXiv171111039T}. The effect of merger-induced metallicity dilution has important consequences for the FMR, because mergers are expected to produce a low-metallicity tail in the residuals around the FMR. This result has been seen in analysis of the observations of \citet{Gronnow-2015} and is also visible in our simulations.

In the present paper we have focused on the effect of gas flows in the central parts of galaxies. Several questions which can be addressed in future research arise from this work. It would for example be interesting to study gas flows by using \emph{Lagrangian gas tracer particles} \citep[for a study of gas tracers within our simulation framework, see][]{2013MNRAS.435.1426G} to quantify how the gas is redistributed within a galaxy in a merger. This would specifically make it possible to estimate to what extent gas is moved between the core of a galaxy, the disk and the circumgalactic medium. Furthermore, it would be possible to study gas accretion from outside the galaxy in detail. In our current set of merger simulations we have not included such gas tracer particles, but in future simulations we plan to do so.

%==================================================================================================
\section{Conclusion}\label{Conclusion}
%==================================================================================================

In this paper we study cosmological simulations of \emph{wet mergers} of main progenitor galaxies with masses in the range $5\times 10^{9} {\rm M}_\odot<M_*<2\times 10^{11}{\rm M}_\odot$ from the Auriga simulation suite. Our sample of mergers is much larger than what has been typically used in studies of idealised merger simulations, and with our full cosmological setup we furthermore have more realistic structure, encounter orbits and environment of our galaxies. Our main results are:

\begin{itemize}
 \item We confirm the result from idealised simulations that strong starbursts occur in major mergers with a mass-ratio larger than $\mu > 1/3$. For minor mergers with $1/10 < \mu <1/3$ there is typically a smaller, but still clearly visible SFR enhancement. Exceptions do, however, occur when several galaxies are involved in the minor merger; in this case the SFR can also increase by an order of magnitude. We note that this scenario, which happens at least occasionally for galaxies in the real Universe, has not been properly explored in previous generations of idealised merger simulations.
 
 \item An increased merger ratio also causes a shorter gas consumption timescale, i.e.~more bursty star formation. We furthermore find that a low-metallicity epoch is typically associated with a merger. The magnitude of the metallicity dilution is strongest for major mergers. Our cosmological simulations thus confirm the occurrence of metallicity dilution, which has previously been seen in observations of merging galaxies and in idealised merger simulations. We consider the period of metallicity dilution as an equally characteristic feature of major mergers as the increased SFR and decreased gas consumption timescale.
 
 \item During the epoch of metallicity dilution mergers have lower metallicity than predicted by the \emph{fundamental metallicity relation} (FMR). This creates a tail of low-metallicity galaxies offset from the FMR. Consistent with findings for observed galaxies the residuals around the FMR indicate that there are more galaxies with low metallicity than predicted by a Gaussian distribution of the residuals. At least part of this effect can be attributed to galaxy mergers.  

\end{itemize}

%==================================================================================================
\section*{Acknowledgements}
%==================================================================================================

We thank Lise Christensen, Kristian Ehlert, Kristian Finlator, Sara Ellison, Maan Hani, Jorge Moreno, Paul Torrey and Mark Vogelsberger for ideas and discussions. We thank the anonymous referee for the constructive and helpful report that helped improve the content of the paper. SB acknowledges support from the International Max-Planck Research School for Astronomy and Cosmic Physics of Heidelberg (IMPRS-HD) and financial support from the Deutscher Akademischer Austauschdienst (DAAD) through the program Research Grants - Doctoral Programmes in Germany (57129429). VS acknowledges the European Research Council through ERC-StG grant EXAGAL-308037, and the SFB-881 `The Milky Way System' of the German Science Foundation. The authors like to thank the Klaus Tschira Foundation. We have used NASAs ADS Bibliographic Services. Most of the computational analysis was performed with {\tt Python 2.7} and its related tools and libraries, {\tt iPython} \citep{Perez-2007}, {\tt Matplotlib} \citep{Hunter-2007}, {\tt scipy} and {\tt numpy} \citep{Van-2011}.

\bibliographystyle{mn2e}

\end{document}